\documentclass[12pt,letterpaper]{article}
\usepackage{amsmath,amssymb}
\usepackage{graphicx,color}

\usepackage[bf]{caption}

\newcommand{\rr}{\mathbb{R}}

\newcommand{\be}{\begin{equation}}
\newcommand{\ee}{\end{equation}}
\newcommand{\ba}{\begin{aligned}}
\newcommand{\ea}{\end{aligned}}
\newcommand{\ben}{\begin{displaymath}}
\newcommand{\een}{\end{displaymath}}
\newcommand{\bea}{\begin{eqnarray}}
\newcommand{\eea}{\end{eqnarray}}
\newcommand{\bean}{\begin{eqnarray*}}
\newcommand{\eean}{\end{eqnarray*}}

\newcommand{\p}{\partial}

\usepackage{amsmath,amssymb}
\usepackage{graphicx,color}

\def\l {\lambda}
\def\th {\theta}
\def\a {\alpha}

\def\b {\beta}

\def\g {\gamma}

\def\e {\epsilon}
\def\s {\sigma}
\def\e {\epsilon}

\def\m{\mu}
\def\n{\nu}

\def\o{\omega}

\definecolor{green}{rgb}{0,0.5,0}

\def\p{\partial}

\usepackage{epsfig}

\long\def\symbolfootnote[#1]#2{\begingroup
\def\thefootnote{\fnsymbol{footnote}}\footnote[#1]{#2}\endgroup}

\begin{document}

\begin{titlepage}

\begin{center}

{\Large \bf On particle type string solutions in $\mbox{AdS}_3\times\mbox{S}^3$}

\vspace{7mm}

{George Jorjadze,$^{a}~$
Zurab Kepuladze,$^b~$
Luka Megrelidze,$^b~$
}\\
[7mm]

{\it${}^a$Razmadze Mathematical Institute and Free University of Tbilisi,}\\
{\it Bedia Str., 0183, Tbilisi, Georgia}\\[3mm]
{\it${}^b$Ilia State University,}\\
{\it K. Cholokashvili Ave 3/5, 0162,
Tbilisi, Georgia}

\vspace{15pt}

\end{center}

\vspace{40pt}

\centerline{{\bf{Abstract}}}
\vspace*{5mm}
\noindent
The $\mbox{AdS}_3 \times\mbox{S}^3$ string dynamics is described in a conformal gauge
using the $\mbox{SL}(2,\rr)$ and $\mbox{SU}(2)$ group variables as the target space coordinates.
A subclass of string surfaces with constant induced metric tensor on both
$\mbox{AdS}_3$ and $\mbox{S}^3$ projections is considered.
The general solution of string equations on this subclass is presented and the corresponding conserved charges
related to the isometry transformations are calculated.
The subclass of solutions is characterized by a finite number of parameters.
The Poisson bracket structure on the space of parameters is calculated,
its connection to the particle dynamics in $\mbox{SL}(2,\rr)\times\mbox{SU}(2)$ is analyzed
and a possible way of quantization is discussed.

\vspace{15pt}
\end{titlepage}

\newpage


\vspace{5mm}

\subsection*{Introduction}

The AdS/CFT correspondence \cite{Maldacena:1997re} is one of the most fruitful research topic
in modern theoretical and mathematical physics of the last decade.
The correspondence is usually realized by
a set of mapping rules for certain quantities of
the partner theories, which are ${\cal N}=4$
supersymmetric Yang-Mills gauge theory in four-dimensional Minkowski
space from one side and $\mbox{AdS}_5\times \mbox{S}^5$ superstring theory from the other.
One of such rules is the map of conformal scaling
dimensions of composite operators of the gauge theory to the energy spectrum of
certain string configurations \cite{Berenstein:2002jq}. After the discovery of an
integrable structure behind the spectral problem of scaling
dimensions \cite{Minahan:2002ve} and the Lax pair formulation of the $\mbox{AdS}_5\times \mbox{S}^5$
string dynamics \cite{Bena:2003wd},
the issue of integrability became the main research line in AdS/CFT \cite{Beisert:2010jr}.

In the present paper we study string dynamics in the $\mbox{AdS}_3 \times \mbox{S}^3$ background,
which can be treated as a subspace of $AdS_5\times S^5$. The paper is a natural continuation of
the work done in \cite{Dorn:2009hs} and \cite{Dorn:2010xt}, where the authors studied spacelike string configurations
in $\mbox{AdS}_3 \times \mbox{S}^3$ with null polygons at the $\mbox{AdS}$ boundary.
The motivation of that work was the analysis of gluon scattering amplitudes at strong
coupling given by a regularized area of string surfaces \cite{Alday:2007hr,aldmald}.

The integration methods used in \cite{Dorn:2009hs, Dorn:2010xt} can be generalized for dynamical strings
replacing the holomorphic structure of Euclidean surfaces with the chiral structure of Lorentzian worldsheets.
An additional new point is the periodicity condition, which has to be imposed for closed string dynamics.

The aim of the work we are starting here is to quantize $\mbox{AdS}_3 \times \mbox{S}^3$ string dynamics and
to investigate its energy spectrum.
This is a hard problem, in general, and in the present paper we restrict ourselves to a subclass of string solutions,
with similar characteristics as in \cite{Dorn:2009hs, Dorn:2010xt}.
In terms of invariant geometrical quantities these are intrinsically flat surfaces,
with constant mean curvatures on both $\mbox{AdS}_3$ and $\mbox{S}^3$ projections.
These restrictions provide a finite
dimensional mechanical system like a particle in $\mbox{AdS}_3 \times \mbox{S}^3$. However, in contrast to a particle,
our string solutions are characterized by winding numbers and they have additional degrees of freedom.

The fact that $\mbox{AdS}_3$ and $\mbox{S}^3$
spaces can be treated as group manifolds, simplifies the analysis of integrability on the basis of
the left and right symmetry transformations. However, for a physical interpretation of results, usually it is more convenient
to use embedding coordinates of $\mbox{AdS}_3$ and $\mbox{S}^3$. Therefore, we apply
both target space coordinates in the text.
Due to the additional freedom mentioned above, the left and right Casimir numbers, in general, are different.
This asymmetry, which is absent for the particle dynamics, has to be realized on the quantum level by a special representation
of the $\mbox{SL}(2,\rr)\times \mbox{SL}(2,\rr)$ and $\mbox{SU}(2)\times \mbox{SU}(2)$ symmetries.

The outline of the paper is the following:
we describe the $\mbox{AdS}_3 \times\mbox{S}^3$ string dynamics
in terms $\mbox{SL}(2,\rr)$ and $\mbox{SU}(2)$ target space variables and conformal worldsheet coordinates.
Components of the metric tensors on the $\mbox{AdS}_3$  and $\mbox{S}^3$ projections are simplified by
turning the chiral and antichiral ones to constants, as in the Pohlmeyer reduction \cite{pohlred}.
We then consider the subclass of worldsheets, which have the remaining components of the metric tensor  also 
constant on both $\mbox{AdS}_3$ and $\mbox{S}^3$ parts.
This subclass is exactly integrable and the corresponding string solutions are characterized by
a finite number of parameters. Among the parameters are four integers which describe different
topological sectors of string configuration.
We calculate the conserved charges related to the isometry transformations and reduce the symplectic structure
of the system to the subspace of solutions.
To simplify the analysis of the physical phase space, we choose a topological sector of the solutions
characterized by one winding number around a cylinder and a torus, which describe string
configurations in $\mbox{AdS}_3$ and $\mbox{S}^3$ projections, respectively.
Finally, we compare the obtained string configurations to the particle dynamics in $\mbox{AdS}_3 \times\mbox{S}^3$ and discuss
a possible way of quantization. Some technical details are given in the Appendix.

\subsection*{$\mbox{AdS}_3$ and $\mbox{S}^3$ as group manifolds}

The $\mbox{AdS}_3$ and $\mbox{S}^3$ spaces are realized as the $\mbox{SL}(2,\rr)$ and $\mbox{SU}(2)$
group manifolds, respectively, via
\begin{equation}\label{g=Y}
  g=\left( \begin{array}{cr}
 Y^{0'}+Y^2 &Y^1+Y^0\\Y^1-Y^0& Y^{0'}-Y^2
 \end{array}\right)~, ~~~~~ h=\left( \begin{array}{cr}
 ~~X^4+iX^3 &X^2+iX^1\\-X^2+iX^1&X^4-iX^3
 \end{array}\right)~.
\end{equation}
Here $(Y^{0'},Y^0,Y^1,Y^2)$ are coordinates of
the embedding space $\rr^{2,2}$ and the equation for the hyperboloid
\begin{equation}\label{YY=-1}
Y\cdot Y\equiv -Y_{0'}^2-Y_0^2+Y_1^2+Y_2^2=-1~,
\end{equation}
which defines the $\mbox{AdS}_3$ space,
is equivalent to $g\in \mbox{SL}(2,\rr)$.
Similarly, the equation
for $\mbox{S}^3$ embedded in $\rr^4$
\begin{equation}\label{XX=1}
X\cdot X\equiv X_1^2+X_2^2+X_3^2+X_4^2=1
\end{equation}
is equivalent to $h\in \mbox{SU}(2)$.

We use the following basis in $\mathfrak{sl}(2,\rr)$
\begin{equation}\label{su(1,1) basis}
  {\bf{t}}_0=\left( \begin{array}{cr}
  ~0&1\\-1&0 \end{array}\right)~,~~~~
   {\bf{t}}_1=\left( \begin{array}{cr}
  0&~1\\1&~0 \end{array}\right)~,~~~~
 {\bf{t}}_2=\left( \begin{array}{cr}
  1&~0\\0&-1 \end{array}\right)~.
\end{equation}
These three matrices ${\bf t}_\mu$ $(\mu=0,1,2)$ satisfy the relations
\begin{equation}\label{tt=}
{\bf{t}}_\mu\,{\bf{t}}_\nu=\eta_{\mu\nu}\,{\bf I}+\epsilon_{\mu\nu}\,^\rho\,
{\bf{t}}_\rho~,
\end{equation}
where
$\eta_{\mu\nu}=\mbox{diag}(-1,1,1)$ and
$\epsilon_{\mu\nu\rho}$ is the Levi-Civita tensor with
$\epsilon_{012}=1$. The inner product defined by
$\langle\, {\bf t}_\mu\,{\bf t}_\nu\,\rangle\equiv\frac{1}{2}\,
\mbox{tr}({\bf t}_\mu\,{\bf t}_\nu)
=\eta_{\mu\nu}$
provides the isometry between $\mathfrak{sl}(2,\rr)$ and 3d Minkowski space.

A similar basis in $\mathfrak{su}(2)$ is given by
the anti-hermitian matrices ${\bf s}_n=i\boldsymbol{\sigma}_n$ $(n=1,2,3)$,
where $\boldsymbol{\sigma}_n$ are the Pauli matrices
($\boldsymbol{\sigma}_1={\bf t}_1,$
$\,\boldsymbol{\sigma}_2=-i{\bf t}_0$, $\,\boldsymbol{\sigma}_3={\bf t}_2$),
and they form the
algebra
\begin{equation}\label{ss=}
{\bf s}_m\,{\bf s}_n=-\delta_{mn}\,{\bf I}-\epsilon_{mnl}\,{\bf s}_l~.
\end{equation}
The inner product is introduced by a similarly normalized trace,
but with the negative sign $\langle{\bf s}_m\,{\bf s}_n\rangle
\equiv -\frac{1}{2}\,\mbox{tr}({\bf s}_m\,{\bf s}_n)=\delta_{mn}$. That
provides the isometry of $\mathfrak{su}(2)$ with $\rr^3$.

Two definitions in \eqref{g=Y} can be written as
$g=Y^{0'}\,{\bf I}+Y^\mu\,{\bf t}_\mu,$ $~h=X_4\,{\bf I}+X_n\,{\bf s}_n$ and the corresponding
inverse group elements are
 $g^{-1}=Y^{0'}\,{\bf I}-Y^\mu\,{\bf t}_\mu,~$
$h^{-1}=X_4\,{\bf I}-X_n\,{\bf s}_n$.
Using now \eqref{tt=} and  \eqref{ss=}, one finds the following relations between the
metrics on these spaces
\begin{equation}\label{dg=dY,dh=dX}
dY\cdot dY=\langle\, (g^{-1}\,dg)\,(g^{-1}\,d g)\rangle~,~~~~~~
dX\cdot dX=\langle\,(h^{-1}\,dh)\,(h^{-1}\,d h)\rangle~.
\end{equation}
These relations allow to write
the $\mbox{AdS}_3 \times
\mbox{S}^3$ string action in terms of the group variables.

\subsection*{String description in terms of group variables}

According to \eqref{dg=dY,dh=dX}, the components of the induced metric tensors on the $\mbox{AdS}_3$ and
$\mbox{S}^3$ projections can be written as\footnote{In this paper
(as in \cite{Dorn:2009hs, Dorn:2010xt}) the
index $s$ is used for some variables of the spherical part to distinguish
them from similar variables of the AdS part.}
\be\label{induced metric A-S}
f_{ab}=\langle\,\left(g^{-1}\,\p_a g\right)
\left(g^{-1}\,\p_b g\right)\,\rangle~,\quad f^s_{ab}=\langle\,\left(h^{-1}\,\p_a h\right)
\left(h^{-1}\,\p_b h\right)\,\rangle~,
\ee
where we use the covariant notation $\p_a=\p_{\xi^a}$ $(a=0,1)$, $(\xi^0,\xi^1)=(\tau,\sigma)$.

A timelike surface in $\mbox{AdS}_3 \times
\mbox{S}^3$ can be parameterized by conformal worldsheet coordinates $z=\tau+\s,~\bar{z}=\tau-\s$ and one
gets a pair of worldsheet fields $g(z,\bar z)\in \mbox{SL}(2,\rr)$ and $h(z,\bar z)\in \mbox{SU}(2)$.
With the notation $\p = \frac{1}{2}(\p_\tau+\p_\s)$, $\,\bar{\p}= \frac{1}{2}(\p_\tau- \p_\s)$,
the conformal gauge conditions take the form
\begin{equation}\label{conformal gauge}
\langle\, \left(g^{-1}\,\partial g\right)^2\,\rangle
+\langle\, \left(h^{-1}\,\partial h\,\right)^2\,\rangle=0=\langle
\left(g^{-1}\,\bar\partial g\,\right)^2\,\rangle+
\langle\,\left(h^{-1}\,\bar\partial h\right)^2\,\rangle~.
\end{equation}

We consider a closed string with periodic ($\s\in \mbox{S}^1$) boundary conditions. Its action in the
gauge \eqref{conformal gauge} is given by
\begin{equation}\label{action}
S=\frac{\sqrt\lambda}{\pi}\int \mbox{d}\tau\int_0^{2\pi}\mbox{d}\sigma ~
[\langle\, (g^{-1}\,\partial
g)\,(g^{-1}\,\bar\partial g)\rangle+
\langle\, (h^{-1}\,\partial
h)\,\,(h^{-1}\,\bar\partial h)\rangle]~,
\end{equation}
where $\lambda$ is a coupling constant, and the equations of motion become
\begin{equation}\label{eq of motions}
\partial\left(g^{-1}\,\bar\partial g\right)+
\bar\partial\left(g^{-1}\,\partial g\right)=0~,\quad \quad \quad
\partial\left(h^{-1}\,\bar\partial h\right)+
\bar\partial\left(h^{-1}\,\partial h\right)=0~.
\end{equation}
These equations provide the chirality conditions
\begin{equation}\label{holomorphicity}
\bar\partial\langle\,\left(g^{-1}\,\partial g\right)^2\,\rangle
=0=\partial\langle \left(g^{-1}\,\bar\partial g\right)^2\,\rangle~,
\quad \bar\partial\langle\,\left( h^{-1}\,\partial h\right)^2\rangle
=0=\partial\langle \left(h^{-1}\,\bar\partial h\right)^2\,\rangle~.
\end{equation}
Using then the freedom of conformal transformations one can map the chiral
$\langle\,\left( h^{-1}\,\partial h\right)^2\rangle$
and antichiral $\langle \left(h^{-1}\,\bar\partial h\right)^2\,\rangle$ components of the metric tensor
on $\mbox{S}^3$ to positive constants.
As a result, from \eqref{conformal gauge} we get the gauge fixing conditions
\begin{equation}\label{gauge fixing}
\langle\,\left(g^{-1}\,\partial g\right)^2\,\rangle
=-\mu^2=-\langle \left(h^{-1}\,\partial h\right)^2\,\rangle~,\quad \quad
\langle\,\left(g^{-1}\,\bar\partial g \right)^2\rangle
=-\bar\m^2=-\langle \left(h^{-1}\,\bar\partial h\right)^2\,\rangle~,
\end{equation}
with constant $\m$ and $\bar\m$.
These constants become dynamical parameters of string solutions like zero modes on a cylinder.

After imposing the gauge fixing conditions \eqref{gauge fixing}, the ${z\bar z}$ component of the metric tensor still remains
arbitrary and due to \eqref{gauge fixing}
its $\mbox{SL}(2,\rr)$ and $\mbox{SU}(2)$ parts can be written as
\begin{equation}\label{z-barz}
\langle\,g^{-1}\,\partial g\,g^{-1}\,\bar\partial g\rangle
=-\m\bar\m \cosh \a~, \quad\quad \langle h^{-1}\,\partial h
\,h^{-1}\,\bar\partial h\rangle=\m\bar\m\,\cos\b~,
\end{equation}
where $\a$ and $\b$ are worldsheet fields.

In the next sections we consider string solutions with constant $\a$ and $\b$. They are
characterized by a finite number of parameters. Note that a constant induced metric tensor on a cylindrical worldsheet
is invariant under translations of $(\tau,\,\s)$ coordinates and this
freedom can be used to reduced the number of parameters by two.
Finally, the obtained dynamical system becomes similar to a
particle in $\mbox{SL}(2,\rr) \times\mbox{SU}(2)$. However, there is also an essential difference,
which has to be taken into account in quantization.

\subsection*{Particle type solutions}

In $(\tau, \sigma)$ coordinates the equations of motion \eqref{eq of motions} takes the form
\be\label{eq of motions in tau-sigma}
\p_\tau(g^{-1}\p_\tau g)-\p_\s(g^{-1}\p_\s g)=0~,\quad\quad \p_\tau(h^{-1}\p_\tau h)-\p_\s(h^{-1}\p_\s h)=0~,
\ee
and the metric tensors \eqref{induced metric A-S} become
\bea\nonumber
&&f_{ab}=\left(\begin{array}{cc}
-2\bar\m\m\cosh\a-\bar\m^2-\m^2& \bar\m^2-\m^2\\
              \bar\m^2-\m^2 & 2\bar\m\m\cosh\a-\bar\m^2-\m^2
            \end{array}\right)~,\\\label{tau-sigma-mertics}
\\ \nonumber
&&f^s_{ab}=\left(\begin{array}{cc}
\bar\m^2+\m^2+2\bar\m\m\cos\b& \m^2-\bar\m^2\\
              \m^2-\bar\m^2 & \bar\m^2+\m^2-2\bar\m\m\cos\b
            \end{array}\right)~.
\eea
The integration of \eqref{eq of motions in tau-sigma} for constant metric tensors \eqref{tau-sigma-mertics}
can be done similarly to the spacelike surfaces with the help of auxiliary linear systems \cite{Dorn:2010xt}.
Constant metric tensors \eqref{tau-sigma-mertics} provide constant coefficients of the linear systems. The integration
is then straightforward and we find the solutions
\begin{equation}\label{solution}
g(\tau,\sigma)=e^{\left(\l\tau +\frac{m}{2}\s\right)\,\hat l}\,g_{_0}\,e^{\left(\rho\tau +
\frac{n}{2}\s\right)\,\hat r}~,
\quad h(\tau,\sigma)=e^{\left(\l_s\tau +\frac{m_s}{2}\s\right)\,\hat l_s}\,h_{_0}\,e^{\left(\rho_s\tau +
\frac{n_s}{2}\s\right)\,\hat r_s}~.
\end{equation}
Here $g_{_0}\in \mbox{SL}(2,\rr)$ and $h_{_0}\in \mbox{SU}(2)$ are constant group elements, $\hat l$ and $\hat r$ are unit timelike elements
of $\mathfrak{sl}(2,\rr)$,
$\hat l_s$ and $\hat r_s$ are unit vectors of $\mathfrak{su}(2)$
\be\label{normalized vectors}
\langle\,\hat l\,\hat l\,\rangle=-1=\langle\,\hat r \, \hat r \,\rangle~,\quad\quad
\langle\,{\hat l}_s\,\hat l_s \,\rangle=1=\langle\,\hat r_s\, \hat r_s\,\rangle~,
\ee
and the other parameters are related to each other by
\be\label{relation between parametrs}
4\l \rho=m n~,\quad\quad \quad 4\l_s \rho_s=m_s n_s~.
\ee
The numbers $m$ and $n$ (as well as $m_s$ and $n_s$) are integers with
a same parity $m-n=2k$ ($m_s-n_s=2k_s$), that provides the periodicity conditions
\be\label{periodicity}
g(\tau,\sigma+2\pi)=g(\tau,\sigma)~, \quad\quad h(\tau,\sigma+2\pi)=h(\tau,\sigma)~.
\ee

The isometry transformations of $\mbox{SL}(2,\rr)$ and $\mbox{SU}(2)$ are given by
the left and right multiplications of the group variables
\be\label{isometry transformations}
g\mapsto g_{_L}\,g\,g_{_R}~,~\quad \quad h\mapsto h_{_L}\,h\,h_{_R}~,
\ee
and they transform the parameters of the solutions \eqref{solution}  by
\bea\nonumber
\hat l\mapsto g_{_L}\,\hat l\, g_{_L}^{-1}~,\quad \hat r\mapsto g_{_R}^{-1}\,\hat r\, g_{_R}~,&
\quad \quad
\hat l_s\mapsto h_{_L}\,\hat l_s\, h_{_L}^{-1}~,\quad \hat r_s\mapsto h_{_R}^{-1}\,\hat r_s\, h_{_R}~,&\\
\label{parameter transform}\\ \nonumber
 g_{_0}\mapsto g_{_L}\, g_{_0}\,g_{_R}~,~~~~~~~~& \quad  h_{_0}\mapsto h_{_L}\, h{_0}\,h_{_R}~,&
\eea
leaving $\l$, $\rho$ and $\l_s$, $\rho_s$  invariant. The integers $m$,  $n$, $m_s$,  $n_s$
are, of course, also invariant. Additional invariants of the isometry transformations are the following parameters
\be\label{theta}
\cosh2\th=-\langle\,\hat l\,g_{_0}\hat r\,g_{_0}^{-1}\,\rangle~,\quad\quad
\cos2\th_s=\langle\,\hat l_s\,h_{_0}\,\hat r_s\,h_{_0}^{-1}\,\rangle~,
\ee
which have an invariant geometrical meaning. Namely,
the mean curvatures of the surfaces in $\mbox{SL}(2,\rr)$
and $\mbox{SU}(2)$ are given by $H=-\coth 2\th$ and $H_s=\cot 2\th_s$, respectively.

Using the isometry transformations \eqref{isometry transformations} one can bring the solutions
\eqref{solution} to the form
\bea\nonumber
g=e^{\th_l\,{\bf t}_0}\,e^{\theta\,{\bf t}_1}\,e^{\th_r\,{\bf t}_0}~,&&\,
\quad\quad  h=e^{\th^s_l\,{\bf s}_3}\,e^{\theta_s\,{\bf s}_2}\,e^{\th^s_r\,{\bf s}_3}~,\\
\label{simple solutions}  \\ \nonumber
\th_l=\l\,\tau +\frac{m}{2}\,\s ~, \quad \th_r=\rho\,\tau+\frac{n}{2}\,\s~,&&\quad\quad
\th^s_l=\l_s\,\tau +\frac{m_s}{2}\,\s ~, \quad \th^s_r=\rho_s\,\tau+\frac{n_s}{2}\,\s~.
\eea
The corresponding $2 \times 2$ matrices are (see Appendix)
\bea\nonumber
&&g=\begin{pmatrix}
  \sinh\th\sin\xi+\cosh\th\cos\eta &
\cosh\th\sin\eta+\sinh\th\cos\xi\\
 \sinh\th\cos\xi-\cosh\th\sin\eta&
\cosh\th\cos\eta-\sinh\th\sin\xi
  \end{pmatrix},\\ \label{simple solutions=} \\ \nonumber
&& h=
\left( \begin{array}{cr}
  \cos\theta_s\,e^{i\xi_s}&\sin\theta_s\,e^{i\eta_s}~\\[0.2cm]
-\sin\theta_s\,e^{-i\eta_s}
& ~\cos\theta_s\,e^{-i\xi_s}~\end{array}\right)~,
\eea
with $\eta=\th_l+\th_r,~$ $\xi=\th_l-\th_r,~$ $\eta_s=\th^s_l-\th^s_r,~$ $\xi_s=\th^s_l+\th^s_r$. The
embedding coordinates of the $\mbox{AdS}_3$ and $\mbox{S}^3$ spaces then become
\bea\nonumber
Y^{0'}=\cosh\th\cos\eta~,\quad~
Y^0=\cosh\th\sin\eta~,\quad~
Y^1=\sinh\th\cos\xi~,\quad
Y^2=\sinh\th\sin\xi~,
\\ \label{AdS-S solutions}\\\nonumber
X_1=\sin\theta_s\,\sin\eta_s~,\quad
X_2=\sin\theta_s\,\cos\eta_s~,\quad
X_3=\cos\theta_s\,\sin\xi_s~,\quad
X_4=\cos\theta_s\,\cos\xi_s~.
\eea
These surfaces represent a tube in $\mbox{AdS}_3$ and a torus in $\mbox{S}^3$ (see Fig.\,1).
Thus, the string is located around the `center' of $\mbox{AdS}_3$ like a static particle in a rest frame.
If one makes a boost transformation, string will oscillate around the center like a massive particle in AdS.

\begin{figure}
\centering{
\begin{tabular}{ccc}
\includegraphics[height=5cm]{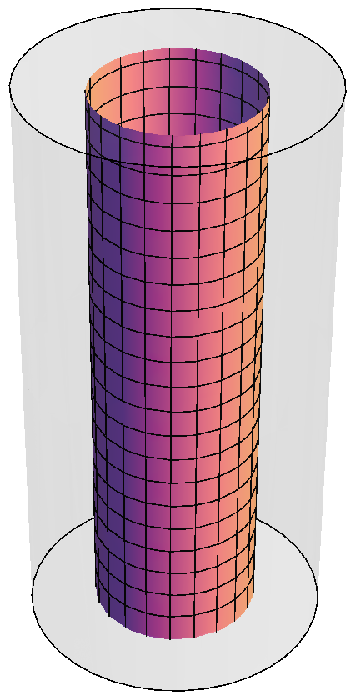}&\hspace{2cm}&\includegraphics[height=4cm]{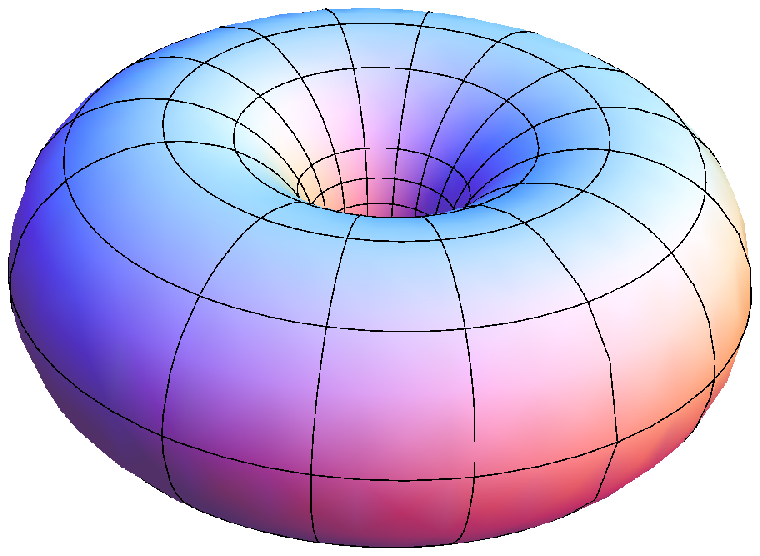}\\
(a)&&(b)
\end{tabular}}
\caption{The plot (a) here corresponds to the AdS projection given by the first line of eq. \eqref{AdS-S solutions} and the plot (b) to the spherical one.}
\label{s1s3}
\end{figure}

To understand the physical characteristics of the solutions \eqref{solution},
we introduce the conserved currents related to the isometry transformations \eqref{isometry transformations}.
The Lie algebra valued
currents in the $\mbox{SL}(2,\rr)$ sector are given by $L_a=\p_ag\,g^{-1},\,\,$ $R_a=g^{-1}\,\p_ag$
and inserting here the solution \eqref{solution}, we find
\bea\nonumber
&L_\tau=\l\,\hat l+\rho\,e^{\th_l\,\hat l}\,g_{_0}\,\hat r\,g_{_0}^{-1}\,e^{-\th_l\,\hat l} ~,
\quad &R_\tau=\l\,e^{-\th_r\,\hat r}\,g_{_0}^{-1}\hat l\,g_{_0}\,e^{\th_r\,\hat r}+\rho\,\hat r\\
\label{currents=} \\ \nonumber
&L_\s=\frac{m}{2}\,\hat l+\frac{n}{2}\,e^{\th_l\,\hat l}\,g_{_0}\,\hat r\,g_{_0}^{-1}\,\e^{-\th_l\,\hat l} ~,
\quad &R_\s=\frac{m}{2}\,e^{-\th_r\,\hat r}\,g_{_0}^{-1}\hat l\,g_{_0}\,e^{\th_r\,\hat r}+\frac{n}{2}\,\hat r~.
\eea

The equation of motion \eqref{eq of motions in tau-sigma} for the $\mbox{SL}(2,\rr)$ part is equivalent
to the current conservation law $\p_\tau R_\tau-\p_\s R_\s=0=\p_\tau L_\tau-\p_\s L_\s$, which is
simply fulfilled by \eqref{relation between parametrs}.

The induced metric tensor on the $\mbox{SL}(2,\rr)$ part can be written as
\be\label{metric-curents}
f_{ab}=\langle\,L_a\,L_b\,\rangle=\langle\,R_a\,R_b\,\rangle~,
\ee
and comparing it with \eqref{tau-sigma-mertics} we obtain the equations
\bea\nonumber
\l^2+\rho^2+2\l\,\rho\cosh 2\th=\bar\m^2+\m^2+2\bar\m\m\cosh\a~,&\\ \label{metric-currents=}
\frac{1}{4}(m^2+n^2+2m n\cosh2\th)=\bar\m^2+\m^2-2\bar\m\m\cosh\a~,&\\ \nonumber
\frac{1}{2}[\l m+\rho n+(\l n+\rho m)\cosh2\th]=\m^2-\bar\m^2~,&
\eea
which connect the parameters of the solution to the components of the induced metric tensor.

The $\mbox{SU}(2)$ conserved currents have the same form
$L^s_a=\p_ah\,h^{-1},\,\,$ $R^s_a=h^{-1}\,\p_ah$
and from \eqref{tau-sigma-mertics} and \eqref{solution} we find the relations similar to \eqref{metric-currents=}
\bea\nonumber
\l_s^2+\rho_s^2+2\l_s\,\rho_s\cos2\th_s=\bar\m^2+\m^2+2\bar\m\m\cos\b~,&\\ \label{S-metric-currents=}
\frac{1}{4}(m_s^2+n_s^2+2m_s n_s\cos2\th_s)=\bar\m^2+\m^2-2\bar\m\m\cos\b~,&\\ \nonumber
\frac{1}{2}[\l_s m_s+\rho_s n_s+(\l_s n_s+\rho_s m_s)\cos2\th_s]=\m^2-\bar\m^2~.&
\eea

The calculation of the $\mbox{SL}(2,\rr)$ conserved charges
\be\label{charges}
L=\int_0^{2\pi} \frac{\mbox{d}\s}{2\pi}\,\, L_\tau ~,\quad\quad
R=\int_0^{2\pi} \frac{\mbox{d}\s}{2\pi} \,\,R_\tau ~,
\ee
for the solution \eqref{solution} yields (see Appendix)
\be\label{charges=}
L=(\l+\rho\,\cosh2\th)\,\hat l~,\quad\quad
R=(\l\,\cosh2\th+\rho)\,\hat r~,
\ee
and, similarly, for the $\mbox{SU}(2)$ charges one gets
\be\label{S-charges=}
L_s=(\l_s+\rho_s\,\cos2\th_s)\,\hat l_s~,\quad
R_s=(\l_s\,\cos2\th_s+\rho_s)\,\hat r_s~.
\ee

One can solve $\m^2$ and $\bar\m^2$ from the equations \eqref{metric-currents=}
and \eqref{S-metric-currents=} separately and then comparing these solutions one finds
two relations between the invariant
parameters $\l$, $\rho$, $\th$ and their spherical counterparts $\l_s$, $\rho_s$, $\th_s$.
Together with \eqref{relation between parametrs}, we conclude that the space of isometrically
invariant parameters is two dimensional. At this point we relate to each other the parameters
of the $\mbox{AdS}_3$ and $\mbox{S}^3$ spaces.
The joint analysis of the equations \eqref{metric-currents=}-\eqref{S-metric-currents=} for arbitrary
$m$, $n$ and $m_s$, $n_s$ is rather complicated
and, in general, they have no consistent solutions. In this paper we concentrate to the case
$m_s=n_s=-m=n>0$.
The corresponding $\mbox{AdS}_3$ and $\mbox{S}^3$ solutions \eqref{AdS-S solutions} become
\bea\nonumber
Y=\big(\cosh\th\cos E\tau,~\cosh\th\sin E\tau,~
\sinh\th\cos(F\tau-n\s),~\sinh\th\sin(F\tau-n\s)\big)~,
\\ \label{AdS-S simple solutions}\\\nonumber
X=\big(\sin\theta_s\,\sin A\tau,~
\sin\theta_s\,\cos A\tau,~\cos\theta_s\,\sin(B\tau+n\s),~
\cos\theta_s\,\cos(B\tau+n\s)\big)~,
\eea
where $E=\l+\rho$, $F=\l-\rho$, $A=\l_s-\rho_s$ and $B=\l_s+\rho_s$.
Due to \eqref{relation between parametrs}, the new parameters are related by
\be\label{quadratic rlations}
F^2-E^2=n^2~, \qquad \qquad B^2-A^2=n^2~,
\ee
where $n$ is the winding number. For fixed $\tau$,
the string \eqref{AdS-S simple solutions} winds $n$-times around the circles
in the $(Y^1,\, Y^2)$ and $(X^3,X^4)$ planes.
Since the polar angle in the $\left(Y^{0'},\,Y^0\right)$ plane
corresponds to the time variable, the solution \eqref{AdS-S simple solutions} describes
string in a static gauge.

From eq. \eqref{metric-currents=} we find
\bea\nonumber
\m^2=\frac{n^2}{4}\,(f-1)(f+\cosh2\th)~,\quad \bar\m^2=\frac{n^2}{4}\,(f+1)(f-\cosh2\th)~,\\ \label{mu-barmu}
\cosh\a=\frac{e}{\sqrt{e^2-\sinh^22\th}}~,~~~~~~~~~~~~~~~~~~~~~~~~~~~
\eea
where $e$ and $f$ are the rescaled variable $e=E/n$, $f=F/n$, with $f^2-e^2=1$.
Eq. \eqref{S-metric-currents=} similarly provides
\bea\nonumber
\m^2=\frac{n^2}{4}\,(b+1)(b+\cos2\th_s)~,\quad \bar\m^2=\frac{n^2}{4}\,(b-1)(b-\cos2\th_s)~,\\ \label{S-mu-barmu}
\cos\b=\frac{a}{\sqrt{a^2-\sin^2 2\th_s}}~,~~~~~~~~~~~~~~~~~~~~~~~~~~~
\eea
with $a=A/n$ $b=B/n$, $b^2-a^2=1$. Choosing $b$ and $f$ as independent variables on the space of invariants, we find
\be\label{Cos=}
\cosh2\th=b f-b^2+1~,\quad \quad \cos2\th_s=f^2-b f-1~.
\ee

Let us consider the parametrization of $g_{_0}$. It can be written as
\be\label{phi}
g_{_0}=e^{\phi_l\,\hat l}\,e^{-(\g+\th)\hat n}\,e^{\phi_r\,\hat r}~,
\ee
where $\hat n$ is a normalized commutator of the matrixes $\hat l$ and $\hat r$ (see \eqref{hat n}),
$\g$ is the corresponding `angle' variable between them and $\phi_l$, $\phi_r$ are arbitrary parameters.
Eq. \eqref{phi} has the following geometrical interpretation: $\hat n$ is a generator of boosts in the
$(\hat l, \hat r)$ `plane' (see \eqref{e^n}) and by \eqref{tr(e^nr)} one finds the boost
parameter $\a=\g+\th$ to match \eqref{theta}. The angle parameters $\phi_l$ and $\phi_r$ in \eqref{phi}
then describe the freedom that leaves \eqref{theta} invariant.

The parametrization of $h_{_0}$ is obtained in a similar way
\be\label{phis}
h_{_0}=e^{\phi^s_l\,\hat l_s}\,e^{-(\g_s+\th_s)\hat n_s}\,e^{\phi^s_r\,\hat r_s}~,
\ee
where $n_s,$ $\g_s$ and $\phi^s_l$, $\phi^s_r$ have the same geometrical
interpretation in $\mbox{SU}(2)$ as their counterparts in $\mbox{SL}(2,\rr)$.

Now we recall that $(\tau,\,\s)$ coordinates still have a freedom in translations.
Using equations \eqref{phi}, \eqref{phis},
and the form of the solutions \eqref{solution}, one can
reduce the number of angle parameters ($\phi_l$, $\phi_r,$ $\phi^s_l$, $\phi^s_r$) from four to two.
We denote these two parameter by $\varphi_1,$  $\varphi_2$.
One can choose, for example, $\varphi_1=\phi_l=-\phi_r$ and $\varphi_2=\phi^s_l=\phi^s_r$.

Thus, solutions \eqref{solution} are described by four unit vectors ($\hat l$, $\hat r$, $\hat l_s$, $\hat r_s$),
two isometrically invariant parameters ($f$, $b$) and two remaining angle variables ($\varphi_1,$  $\varphi_2$).
Totally, one gets twelve dimensional space of parameters.

To find the Poisson bracket structure on this space, one can calculate
the symplectic form of the system $\o=\mbox{d}\vartheta$ on the space of solutions \eqref{solution} and invert it.
The presymplectic 1-form $\vartheta$ is defined from the string Lagrangian in the first order formalism
\be\label{string 1-form}
\vartheta=\int_0^{2\pi}\frac{\mbox{d}\s}{2\pi}\,\left[\langle R\,g^{-1}\,\mbox{d}g\rangle+
\langle R_s\,h^{-1}\,\mbox{d}h\rangle\right]~.
\ee
Here $R$ and $R_s$ are Lie algebra valued variables, which in the conformal
gauge are associated with $g^{-1}\p_\tau g$ and $h^{-1}\p_\tau h$, respectively.
Before we discuss the reduction of the symplectic form on the string solutions, let us consider particle dynamics
in $\mbox{SL}(2,\rr)\times \mbox{SU}(2)$ and compare it to our system.

Particle trajectories in the $\mbox{SL}(2,\rr)$ sector are
parameterized
either by a pair ($g_{_0}, \,R$) or  ($g_{_0}, \,L$)
\be\label{solution-p}
g(\tau)=e^{L\,\tau}\,g_{_0}=g_{_0}\,e^{R\,\tau}~,
\ee
where $R$ and $L$ are the dynamical integrals for the isometry transformations.\footnote{We use same notations as for string solutions.}
They are related
to each other by the adjoint transformation
\be\label{gR=Lg}
g\,R\,g^{-1}=L~,
\ee
and, therefore, they are on the same coadjoint orbit.

The description of $\mbox{SU}(2)$ sector is similar and the orbits in both cases are defined by the Casimir numbers
\be\label{LL=RR}
\langle\,L\,L\,\rangle=\langle\,R\,R\,\rangle=-m^2~, \quad
\quad \langle\,L_s\,L_s\,\rangle=\langle\,R_s\,R_s\,\rangle=m_s^2~.
\ee
These numbers are related to the particle mass $M$ by the massshell condition
\be\label{massshell}
m^2-m_s^2=M^2~,
\ee
and the dynamical integrals can be written as
\be\label{L,R=}
L=m\,\hat l~, \quad \quad R=m\,\hat r~;\quad \quad L_s=m_s\,\hat l_s~, \quad \quad \quad
R_s=m_s\,\hat r_s~,
\ee
where $\hat l,$  $\hat r$ and  $\hat l_s,$  $\hat r_s$ are unit vectors as for the string solutions (see \eqref{charges=}-\eqref{S-charges=}).

Hamiltonian formulation can be started in $(R,\,g)$, $(R_s,\,h)$ variables
and the presymplectic 1-form $\vartheta=\langle R\,g^{-1}\,\mbox{d}g\,\rangle+\langle R_s\,h^{-1}\,\mbox{d}h\,\rangle$,
as in \eqref{string 1-form}. However, in order to make further comparison with the string solutions,
it is more convenient to use left and right dynamical integrals symmetrically
and express $g$ and $h$ through them.

Let us consider the $\mbox{SL}(2,\rr)$ part.
One can show that eq. \eqref{gR=Lg} defines $g$ up to an angle variable $\varphi$ \cite{Dzhordzhadze:1994nv}
\be\label{g=phi}
g=e^{\th_{_L}\,\hat n_{_L}}\,e^{\varphi\,{\bf t}_0}\,e^{\th_{_R}\,\hat n_{_R}}~.
\ee
Here
\be\label{nL,n_R}
\hat n_{_L}=\frac{[{\bf t}_0,\,\hat l]}{2\sinh2\th_{_L}}~,\quad \quad
\hat n_{_R}=-\frac{[{\bf t}_0,\,\hat r]}{2\sinh2\th_{_R}}~
\ee
are normalized vectors and  $\cosh2\th_{_L}=-\langle\,\hat l\,{\bf t}_0\,\rangle,\,$
$\,\cosh2\th_{_R}=-\langle\,\hat r\,{\bf t}_0\,\rangle$ (see Appendix).

The symplectic form of the AdS part $\o_{_{AdS}}=\mbox{d}\langle R\,g^{-1}\,\mbox{d}g\rangle$
calculated in the coordinates $(\hat l,\hat r, m, \varphi)$ splits into the sum of three terms
\be\label{splitting}
\o_{_{AdS}}=m\,\o_{_L}+m\,\o_{_R}-\mbox{d}m \wedge \mbox{d}\varphi~,
\ee
where
\be\label{AdS 2-form}
\o_{_L}=\frac{\mbox{d}l_2\wedge \mbox{d}l_1}{2l^0}~,\quad\quad \o_{_R}=\frac{\mbox{d}r_1\wedge \mbox{d}r_2}{2r^0}
\ee
are the symplectic forms on the unit $\mbox{SL}(2,\rr)$ coadjoint orbits expressed in terms of the vector components
$l_\m=\langle\,{\bf t}_\m\,\hat l\,\rangle$ and $r_\m=\langle\,{\bf t}_\m\,\hat r\,\rangle$.

The $\mbox{SU}(2)$ part of the symplectic form has a similar structure
\be\label{splitting-s}
\o_{s}=m_s\,\o_{_L}^s+m_s\,\o_{_R}^s+\mbox{d}m_s \wedge \mbox{d}\varphi_s,
\ee
where now $\o_{_L}^s$ and $\o_{_R}^s$ are symplectic forms on the unit $\mbox{SU}(2)$ coadjoint orbits.

Finally, the total symplectic form $\omega=\o_{_{AdS}}+\o_s$ has to be reduced
on the massshell \eqref{massshell}. This reduction leads to a ten
dimensional phase space with the symplectic form
\be\label{particle SF}
\o=m\,\o_{_L}+m\,\o_{_R}+m_s\,\o_{_L}^s+m_s\,\o_{_R}^s+\mbox{d}m_s \wedge
\mbox{d}\left(\varphi_s-\frac{\varphi}{\sqrt{m_s^2+M^2}}\right),
\ee
where $\,m\,=\,\sqrt{M^2+\,m_s^2}\,.\,\,$ The reduced phase space is parameterized by four unit vectors 
($\hat l$, $\hat r,$ $\hat l_s$, $\hat r_s$),
one isometrically invariant parameter $m_s$ and the corresponding angle variable.
The inversion of \eqref{particle SF} provides a Poisson bracket realization of the left-right symmetries
\be\label{PB algebra}
\{L_\m,\,L_\n\}=-2\e_{\m\n}\,^\rho\,L_\rho~, \quad\quad \{R_\m,\,R_\n\}=2\e_{\m\n}\,^\rho\,R_\rho~
\ee
with $L_\m=\langle\,{\bf t}_\m\,L\,\rangle$, $R_\m=\langle\,{\bf t}_\m\,R\,\rangle$ and similarly in the
$\mbox{SU}(2)$ part.

The quantization of the particle dynamics on the basis of the symplectic form \eqref{particle SF} is straightforward and
it reproduces the same spectrum as other quantization schemes \cite{Dorn:2010wt}. Details of the quantization of the particle dynamics
will be presented in a forthcoming paper, which will include the analysis of more general string solutions as well.

At the end of the present paper we return to the $\mbox{SL}(2,\rr)\times\mbox{SU}(2)$ string dynamics.
The calculation of the presymplectic 1-form \eqref{string 1-form} on the space of solutions \eqref{solution} is similar to the calculation
of the conserved charges given in Appendix. The exterior derivative then acts on the space of parameters and
provides the following symplectic form
\be\label{string SF}
\o=m_{_L}\,\o_{_L}+m_{_R}\,\o_{_R}+m^s_{_L}\,\o_{_L}^s+m^s_{_R}\,\o_{_R}^s+
\tilde\o(f,\,b;\,\varphi_1,\,\varphi_2)~.
\ee
Here $m_{_L},\,$ $m_{_R},\,$ $m^s_{_L},\ $ $m^s_{_R}\,$ are the coefficients of the unit vectors in \eqref{charges=}-\eqref{S-charges=}
and they define the Casimir numbers. An essential difference with the particle case is the asymmetry between the
left and right Casimir numbers here. The rest part of the symplectic form \eqref{string SF} given by
$\tilde\o(f,\,b;\,\varphi_1,\,\varphi_2)$ is rather complicated. However, it does not contribute
to the Poisson bracket structure of dynamical integrals. In particular, the
isometrically invariant variables $m_{_L},\,$ $m_{_R},\,$ $m^s_{_L},\ $ $m^s_{_L}\,$
have vanishing Poisson brackets with themselves and with the dynamical integrals.
As it was mentioned above, the space of isometrically invariant variables is two dimensional.
The structure of this space defines the character of coadjoint orbits, that
is crucial for the symmetry group representations.

Quantization based on the symplectic form \eqref{string SF} is in progress.

\vspace{5mm}

\noindent
{\bf Acknowledgments}

\vspace{3mm}

\noindent
We thank Harald Dorn, Chrysostomos Kalousios, Jan Plefka and Sebastian Wuttke  for useful discussions.
The main part of the work has been done at Humboldt University of Berlin during our visits there.
We thank the department of Physics for warm hospitality.

This work has been supported in part by the grant I/84600 from VolkswagenStiftung.

\setcounter{equation}{0}

\def\theequation{A.\arabic{equation}}


\subsection*{Appendix}

Here we present useful formulas for the $\mbox{SU(2)}$ and $\mbox{SL}(2,\rr)$ groups
and sketch some calculations used in the main text.

From the algebras \eqref{tt=} and \eqref{ss=} follow simple exponentiation rules
\bea\nonumber
&e^{\theta\,{\bf t}_0}=\left( \begin{array}{cr}~ \cos\theta&\sin\theta\\
-\sin\theta&\cos\theta
\end{array}\right),\quad \quad
&e^{\theta\,{\bf t}_1}=\left( \begin{array}{cr}~ \cosh\theta&\sinh\theta\\
\sinh\theta&\cosh\theta
\end{array}\right),\\ \label{e^t}\\
&e^{\theta\,{\bf s}_2}=\left( \begin{array}{cr}~ \cos\theta&\sin\theta\\
-\sin\theta&\cos\theta
\end{array}\right)~, \quad \quad
&e^{\th\,\boldsymbol{\bf s}_3}=\left( \begin{array}{cr}~ e^{i\th}&0\\
0&e^{-i\th}
\end{array}\right),
\eea
which gives to the solutions \eqref{simple solutions} a compact matrix form \eqref{simple solutions=}.

Our convention on signatures correspond to the following summation rule
\begin{equation}\label{epsilon epsilon}
\epsilon_{\mu\nu\rho}\epsilon_{\mu'\nu'}\,^\rho=
\eta_{\mu\nu'}\,\eta_{\nu\mu'}-\eta_{\mu\mu'}\,\eta_{\nu\nu'}~.
\end{equation}

The charge $R$  in \eqref{charges} can be written as $R=\lambda \hat J+\rho\,\hat r$,
where $\hat J$ is the integral (see \eqref{currents=})
\be\label{integral}
\hat J=\int_0^{2\pi}\frac{\mbox{d}\s}{2\pi}~ e^{-\th_r\,\hat r}\,\hat l'\,e^{\th_r\,\hat r}~,
\ee
with $\hat l'=g_{_0}^{-1}\,\hat l\,g_{_0}$ and $\th_r$ given by \eqref{simple solutions}.
The exponents $\,e^{\pm\th_r\,\hat r}=\cos\th_r\, {\bf I}\pm\sin\th_r\,\hat r\,\,$ create $\s$ dependent
trigonometric functions in \eqref{integral} and after integration we find $\hat J=\frac{1}{2}\,(\hat l'- \hat r\,\hat l'\,\hat r)$.
By \eqref{tt=} and \eqref{epsilon epsilon} one gets
\be\label{rlr}
\hat r\,\hat l'\,\hat r=\hat l'+2\langle\,\hat r\,\hat l'\,\rangle\,\hat r~,
\ee
and taking into account \eqref{theta}, we obtain
$\hat J=\cosh2\th\,\,\hat r$. This leads to eq. \eqref{charges=} for $R$.
The calculation of other charges is similar.

Let us introduce a normalized commutator of $\hat l$ and $\hat r$
\be\label{hat n}
\hat n=\frac{[\hat l,\hat r]}{2\sinh2\g}~,\quad \mbox{with} \quad  \cosh2\g=-\langle\,\hat l\,\hat r\,\rangle~.
\ee
This matrix satisfies the relations $\,\hat n^2={\bf I},\,$  $\,\hat n\,\hat r=-\hat r\,\hat n\,$
and it generates the boost transformation between $\hat l$ and $\hat r$
\be\label{e^n th}
e^{-\g\hat n}\,\hat r e^{\g\hat n}=e^{-2\g\hat n}\,\hat r=\hat l~.
\ee
The general boost transformation of $\hat r$ is given by
\be\label{e^n}
e^{-\a\hat n}\,\hat r e^{\a\hat n}=\frac{\sinh2(\g-\a)}{\sinh2\g}\,\hat r+\frac{\sinh2\a}{\sinh2\g}\,\hat l~,
\ee
and provides
\be\label{tr(e^nr)}
\langle\,\hat l\,e^{-\a\hat n}\,\hat r e^{\a\hat n}\,\rangle=-\cosh2(\g-\a)~.
\ee
This equation is used in \eqref{phi} to fix the boost parameter $\a$ in $g_{_0}$.

\end{document}